\documentclass[12pt,twoside]{article}
\usepackage{fleqn,espcrc1}

\usepackage{graphicx}
\newcommand{\AmS}{{\protect\the\textfont2
  A\kern-.1667em\lower.5ex\hbox{M}\kern-.125emS}}

\title{Hot fragmentation of nuclei\footnote{
Invited Talk at Seventh International Conference on Nucleus-Nucleus 
Collisions,\\
Strasbourg, France, July 3 - 7, 2000}
}

\author{W. Trautmann\address[GSI]{
	Gesellschaft f\"ur Schwerionenforschung (GSI) \\ 
	D-64291 Darmstadt, Germany}}
\begin{document}

% typeset front matter
\maketitle

\begin{abstract}

Today, we have a variety of reactions at hand that can 
be used to multi-fragment nuclei.
In many of these reactions even several sources of 
fragments can be discerned and characterized.

There is overwhelming evidence that these sources of fragments are hot.
It is already less clear whether heat by itself is sufficient to initiate 
the fragment decay. What causes fragmentation, and when and how are the 
fragments (pre)formed? These questions have remained as much a challenge as 
the complementary class of questions to which they are related: What 
observations derive their significance from the liquid-gas phase behavior 
of extended nuclear matter? And, can we observe a phase transition in 
finite nuclei?

Recent developments, largely coming from complex analyses of data sets 
measured in 4-$\pi$-type experiments as well as from calculations based 
on advanced theoretical concepts, will be discussed.

\end{abstract}

\section{INTRODUCTION}

Fragmentation is a term commonly used to specify a nuclear 
disassembly by force. Hot fragmentation is meant to indicate the most
violent of these processes, following excitations beyond the limits of 
nuclear binding, but still ending with bound nuclear fragments of different
sizes in the final channels. The formation mechanism of these fragments, 
whether they are the remnants of an incomplete destruction or the products
of a condensation ('selforganization', cf. Ref. \cite{bond95a}) out of 
the disordered matter, has continued to be the topic of very active 
research in recent years \cite{more93,hirsch99,rich00}.

A two-stage scenario has proven to be fruitful for the interpretation and 
modelling of hot-fragmentation reactions. It is motivated by the
differences of the wave lengths and time scales governing the entrance 
and exit channels and justified by the remarkable success of statistical
approaches for the second stage \cite{gross90,bond95,radu00}. The 
intermediate states are not necessarily equivalent to hot nuclei but
should be, more generally, viewed as systems of highly excited nuclear 
matter, populating a phase space characterized by global quantities
like mass, charge, energy, density or temperature. 
Molecular-dynamics calculations can now continuously follow the reaction
process from the first encounter to the final disassembly stages
without the need to specifically assume equilibration 
\cite{barz96,goss97,ono99}. In that sense, they present a challenge to
the statistical two-stage picture. On the other hand, the extraction of 
thermodynamical parameters from transport-model calculations has 
demonstrated that a connection to the statistical approach can be 
established \cite{fuchs97}.

There is little doubt that the densities are low and the 
temperatures high in the intermediate state, coinciding with the values 
predicted for the coexistence region of liquid and gaseous nuclear matter.
To search for observable links of the multifragmentation phenomenon 
to the predicted phase transition has 
therefore been a major motivation for many experimental and theoretical 
activities. It is evident from the titles and abstracts submitted to 
this conference that there is growing confidence that signals of this
phase transition are actually being observed.

\begin{figure}[ttb]
\centering
    \includegraphics[width=20pc]{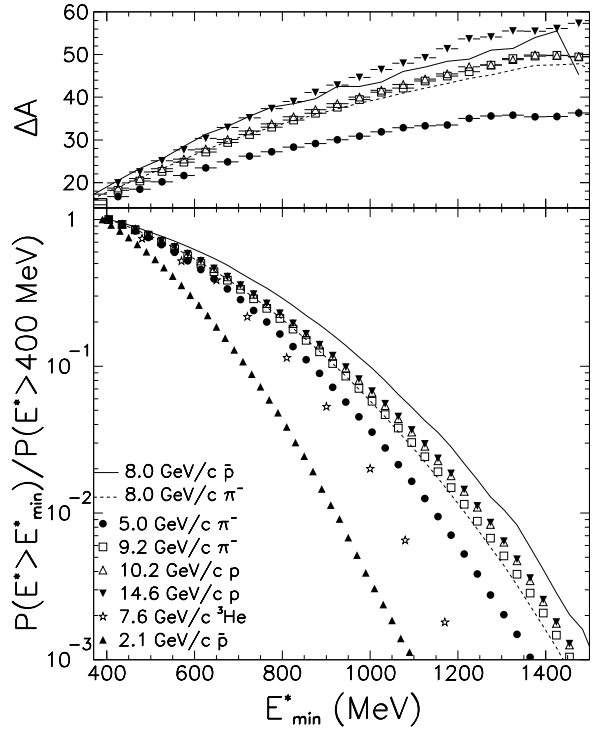} 
\caption{Reactions of energetic hadrons with 
$^{197}$Au targets: average mass loss $\Delta A$ in the fast cascade 
(top) and relative probability for excitation energies 
$E^* > E^*_{\rm min}$ (bottom)
as a function of $E^*_{\rm min}$ (from Ref. \protect\cite{beau99}).
}
\end{figure}

This talk divides into three main parts: a brief review 
of the reactions that are being used to produce hot nuclear systems,
a discussion of the 
caloric curve of nuclei as a potential signature of the liquid-gas phase 
transition in finite systems, and new experimental results for 
the conditions at breakup in spectator reactions. This last 
part will mainly reflect recent activities of the ALADIN collaboration.

\section{USEFUL REACTIONS}

Thermal fragmentation denotes a concept of studying the
breakup of thermally excited nuclear systems formed in collisions of 
relativistic hadrons with heavy target nuclei. It is based on the 
expectation, tested with intranuclear-cascade calculations, that 
light energetic projectiles will generate a statistical (thermal)
disorder without exciting collective modes such as compression, rotation
or shape deformations. The latter are known to break nuclei very 
efficiently \cite{nemeth86,colo97}, 
and thus would mask the nuclear response to the thermal excitation.

The potential and the limits of this approach have been explored by the 
ISiS collaboration in their recent experiments at the AGS in Brookhaven
with a variety of primary and secondary beams including pions and 
antiprotons in the momentum range up to 14.6 GeV/c
\cite{lefort99,beau99,beau00}. The properties of intermediate systems 
produced in these reactions are summarized in Fig. 1.
Target-like residues with several hundreds of MeV excitation 
have the largest cross sections, but excitation energies
exceeding 1 GeV can be reached, even though with rapidly dropping
cross sections. There is also a loss of mass caused by the heating 
mechanism at relativistic energies. Some of the nucleons participating in 
the cascading processes are too energetic to remain part of the 
intermediate system.

\begin{figure}[ttb]
\centering
    \includegraphics[width=24pc]{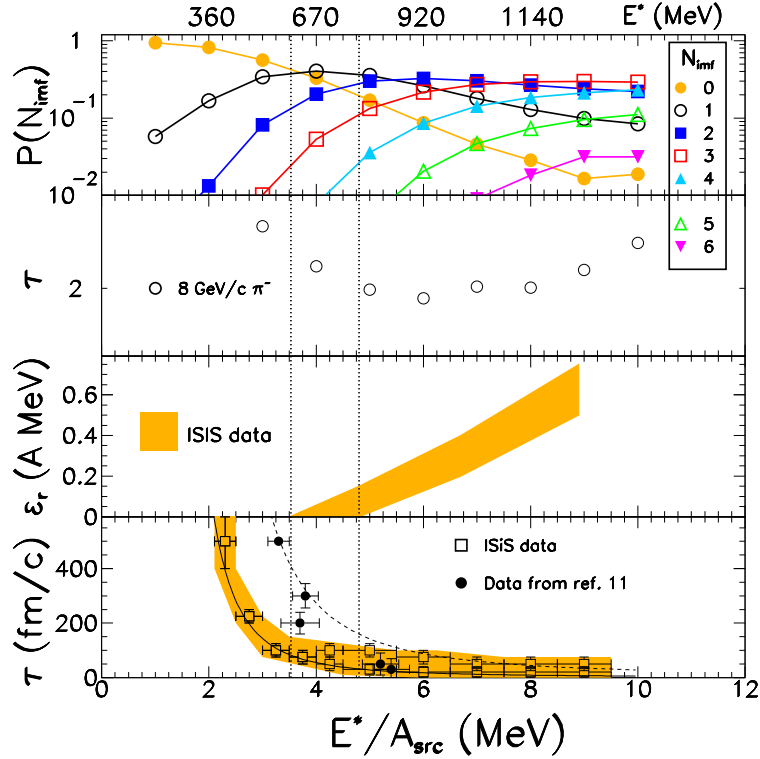} 
\caption{Reactions of energetic hadrons with 
$^{197}$Au targets: distribution of fragment multiplicities
$N_{\rm imf}$, $\tau$  
parameter describing the element distribution, radial flow energy, and
emission time scale (from top to bottom).
Ref. 11 in the bottom panel refers to \protect\cite{dodu98}
(from Refs. \protect\cite{beau99,beau00}).
}
\end{figure}

The multi-fragment channels open up at the higher excitation energies,
as illustrated in Fig. 2. The transition from residue production to hot 
fragmentation, highlighted by the dotted vertical lines, is, most notably,
associated 
with a striking decrease of the emission time scales to values in the 
vicinity of 50 fm/c.

According to Fig. 1, the most efficient projectiles are antiprotons
of high momentum which only are difficult to use because of their low 
abundance in the secondary beam \cite{lefort99}. Some extra heating is 
generated by the pion cascades from the $\bar {\rm p}$ annihilation
which leads to considerable excitation energies already at much lower 
$\bar {\rm p}$ momenta. The reported fragment 
multiplicities, however, are significantly smaller for the less energetic 
antiproton beams even if identical bins of excitation energy are selected
\cite{jahnke99}. It is therefore an open question whether the
excitation energy by itself is the only parameter that governs the decay
properties and fragment production. Similar questions have recently 
been raised by other authors who find fewer fragments in the experiment 
than are predicted by statistical models \cite{sun00,lolly00}. 
While this may be partly
connected to the difficulties inherent with experimentally determining 
the thermalized excitation energy \cite{schuett96}, it is nevertheless 
obvious that the solution to this problem will help us to better understand 
how fragments are formed \cite{odeh00}.

\begin{figure}[ttb]
\centering
\begin{minipage}[c]{.45\textwidth}
   \centering
    \includegraphics[width=15pc]{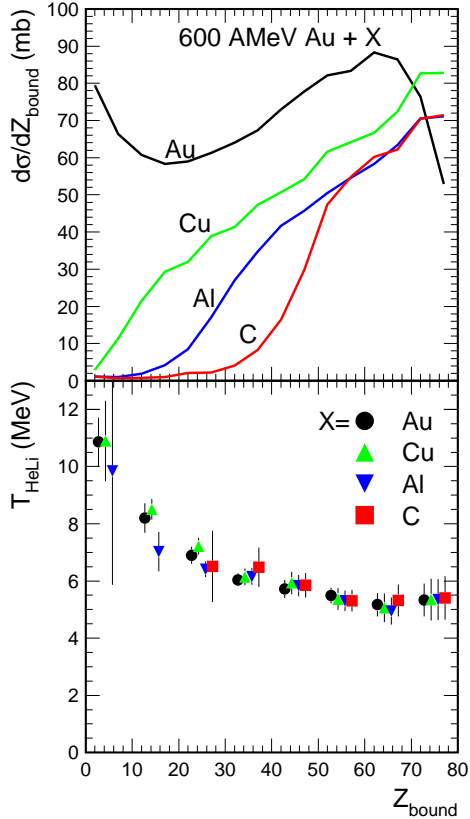} 
\end{minipage}
\begin{minipage}[c]{.45\textwidth}
   \centering
   \caption{Differential cross section ${\rm d}\sigma /{\rm d}Z_{\rm bound}$
(top) and isotope temperature $T_{\rm HeLi}$ (bottom) as a function of
$Z_{\rm bound}$ for the reaction $^{197}$Au on Au, Cu, Al and C targets at
600 AMeV.
Note that the experimental trigger suppresses the very peripheral
collisions at $Z_{\rm bound} \ge$ 65 which have much larger
cross sections than indicated here (from Ref. \protect\cite{odeh99}).
}
\end{minipage}
\end{figure}

The limits of generating excitation energy can be overcome with composite 
projectiles. Systematic sets of data with projectiles of different mass 
and energy have been collected by the FASA \cite{avde98,karn99}, 
EOS \cite{haug98,insol00,haug00} and KEK/HIMAC \cite{tanak95} collaborations, 
and new results are reported in contributions 
to this conference. The continuing rise of 
the cross section for high excitations with increasing mass of the 
collision partner is demonstrated in Fig.~3 for the fragmentation of
$^{197}$Au projectiles studied by the ALADIN collaboration. 
These data are shown as a function of the variable $Z_{\rm bound}$,
representing the sum of the atomic numbers $Z_{\rm i}$ of all projectile 
fragments with $Z_{\rm i} \geq$ 2, which
is inversely correlated with the excitation energy \cite{schuett96}. 
As an example of the $Z_{\rm bound}$ scaling, a prominent feature of spectator 
reactions up to very high energies \cite{cherry95}, 
the isotope temperatures $T_{\rm HeLi}$ are also given in the 
figure. They are close to 6 MeV for the major part of the $Z_{\rm bound}$ 
range, but tend to higher values at low $Z_{\rm bound}$, i.e. at the highest 
excitation energies reached in these reactions. The temperatures depend
only on $Z_{\rm bound}$ but not on the specific target that is used to 
fragment the $^{197}$Au projectile.

Large systems at even higher excitations can be produced in central 
collisions of heavy systems. In this case, the idea of excitation as a
simple heating process has to be abandoned, however. Collective 
modes, compression as well as the directed outward motion of particles and 
clusters from primary collisions generate an explosive pattern, quantified as 
collective radial flow \cite{reis97a}. The production of large clusters 
is rare and inversely correlated with the observed amount of flow.
An extreme case, observed at the AGS, has been reported very recently.
For $^{197}$Au beams of 11.5 GeV/c per nucleon,
centrally colliding with heavy targets, 
the fragment mass yields are steeply exponential with a penalty factor 
of about 50 for each additional mass unit \cite{arm99}. 
The characteristic transition from power law 
to exponential spectra, as radial flow sets in \cite{kunde95}, 
has been reproduced with molecular dynamics 
calculations, very recently e.g. with quantum molecular dynamics
\cite{chika00}, but to fully understand 
the clusterization mechanism in the dynamical environment still remains an 
interesting problem for future research \cite{bond95a,nebau99}. 

Heavy symmetric systems below the threshold of collective radial flow
(about 50 MeV per nucleon) have been extensively studied by the
Miniball/Multics and INDRA collaborations. High-statistics data permit the 
selection of single-source formation which occurs with small cross sections
in central collisions \cite{dago96,rivet98}. 
In more peripheral encounters,
several sources contribute to the fragment yields with a clear enhancement 
in the mid-rapidity domain \cite{lukas97}. The potential mechanisms of 
mid-rapidity emissions, as e.g. neck formation, and the isospin effects
associated with it \cite{demp96,plagnol00} are closely 
connected to several dynamical and statistical
aspects of hot fragmentation, among them the isotopic separation expected 
in the liquid-gas coexistence zone \cite{muell95,chomaz99,xu00}. These topics 
are the subjects of other plenary talks at this conference.

\section{LIQUID-GAS PHASE TRANSITION}

It is commonly accepted that extended nuclear matter should exhibit
a liquid-gas phase transition, following from the Van-der-Waals-like range 
dependence of the nuclear force \cite{muell95,sauer76,jaqa83}.
Sharp discontinuities in the infinite system are expected to broaden
as the system size decreases \cite{campi88,dasgup98}. This, however, does 
not contradict the existence or prevent the
identification of phase transitions in small 
systems with constituent numbers on the nuclear scale \cite{rich00,gross97}.

\begin{figure}
\centering
    \includegraphics[width=32pc]{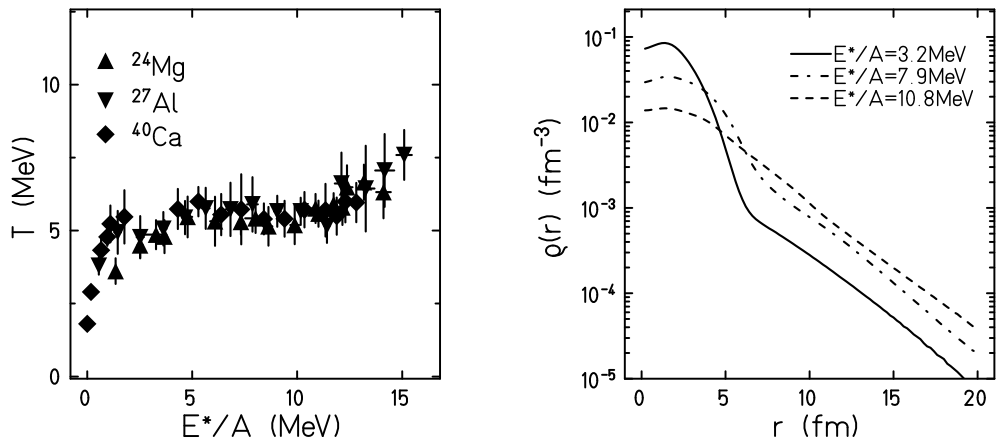} 
\caption{
Results of FMD calculations:
caloric curves of $^{24}$Mg, $^{27}$Al, and $^{40}$Ca 
%at $\hbar \omega$~=~1 MeV 
(left panel) 
and time-averaged radial density distributions of 
$^{24}$Mg at various excitation energies in the coexistence region
(right panel; from Ref. \protect\cite{schnack97}).
}
\label{fig:feld}
\end{figure}

The appropriate experiment for identifying the liquid-gas phase transition 
in finite nuclei has recently been done, theoretically. Fermionic as well
as antisymmetrized molecular dynamics (FMD, AMD) models were used to study 
the equilibrium dynamics of small nuclear systems \cite{schnack97,suga99}.
None of the necessary ingredients were missing in these studies, 
a container to confine the system, a controllable heating technique, 
long propagation times to allow the system to settle into equilibrium, 
and a suitable technique of measuring the temperature. As a remarkable 
result it was shown that the nuclear dynamics, as represented in these 
advanced transport-type models, generate a phase transition without
any further assumptions.

Caloric curves obtained in the FMD study are shown in Fig. 4. The extracted
temperatures exhibit a plateau that extends over about 8 MeV per nucleon 
of excitation energy and has its origin in the coexistence of liquid 
and nuclear phases in the system. The densities reflect the transition from
a liquid phase in equilibrium with its surrounding vapor to the pure vapor
phase. The properties of the system in these asymptotic states were
identified as those of a Fermi liquid and a Fermi gas 
(Van-der-Waals gas in the AMD). Both groups have also demonstrated
that the external conditions, the container which controls the pressure, 
have a strong influence on the properties of the transition. 
Changing the confinement gives the possibility to map out the phase 
diagram. The latent heat of the $^{16}$O system approaches zero at 
$T \approx$ 10 MeV which may be associated with the critical temperature
for that system.

There is no external confinement in the real experiment, 
and the time scales of
hot-fragmentation reactions are rather short. It is therefore even more
surprising that the temperature-energy correlation measured for the
breakup states exhibits such a similar behavior. The data
for $^{197}$Au fragmentation shown in Fig. 5 represent the results for
600~MeV per nucleon \cite{poch95}, 
with small modifications due to additional
experimental information and corrections, and the results for 1000 MeV per 
nucleon obtained more recently \cite{muell99}. The temperature
of the transition region is close to that obtained with the
dynamical \cite{schnack97} or statistical models 
\cite{radu00,xi97,bond98} and does not change with the
bombarding energy. In contrast to it, the energy  
associated with the spectator source increases by, on the average,
30\% over the range 600 to 1000 MeV per nucleon, a
behavior inconsistent with the universality of the
spectator decay that so clearly appears in other variables \cite{schuett96}.
It is caused by the energy dependence
of the mean kinetic energies of nucleons in the spectator frame and 
most likely indicates that contributions from the early stages of the 
reaction have been included in the calorimetry for the spectator source. 
As a consequence,
the apparent latent heat deduced even for the lower energy of 
600 MeV per nucleon should be considered as an upper limit.

\begin{figure}[ttb]
\centering
    \includegraphics[width=20pc]{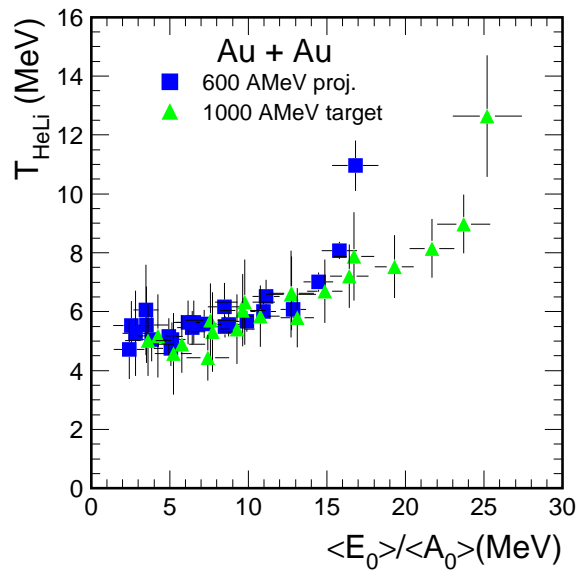} 
\caption{Caloric curve for spectator decays in $^{197}$Au on $^{197}$Au 
collisions. The temperatures were determined 
from helium and lithium isotope ratios of projectile fragments at 600~AMeV 
and of target fragments at 1000~AMeV
(from Ref. \protect\cite{odeh99}).
}
\end{figure}

Breakup temperatures and energy contents of the fragmenting system 
were measured and correlated by several other groups for a variety of 
different reactions \cite{haug98,ma97,kwiat98,dago99,cibor00}. 
The basic methods were identical with some variation in the 
approximations that had to be made. Differences exist, e.g. in 
whether and how the effects of sequential decays on the temperature were
taken into account, whether the neutron multiplicities and kinetic 
energies were measured or estimated, and whether and how preequilibrium 
components were identified and explicitly excluded. The last point is 
part of the bigger problem of identifying and properly selecting the 
fragmenting source which is crucial. 

The resulting caloric curves have several features in common but 
are considerably different in detail. 
A deviation from the behavior of a Fermi liquid is observed for all 
reactions but at different temperatures between 5 and 7 MeV. The slopes
are somewhat different and the upbend at high excitations is only 
seen in the $^{197}$Au spectator decay (Fig. 5). It is not observed
for the other reactions which, however, 
do not easily lead to comparable excitation energies (Fig. 3). 

Apart from the experimental differences and imperfections, it is the
transient nature of the reaction process which most likely 
prevents a single universal curve to emerge from these studies.
To the extent that the equilibrated breakup state is 
an idealization, measured fragment yields represent integrals over finite 
emission times \cite{cibor00,xi98,viola99}, with pre-breakup and post-breakup 
contributions varying among different reactions. Furthermore, 
the expansion dynamics may generate transient pressures that are different 
for different types of hot-fragmentation reactions. The prominent role
of the pressure, however, is known from the model experiments
\cite{schnack97,suga99}.

\section{PARAMETERS OF THE BREAKUP STATE}

Temperatures and excitation energies are only two of the quantities of 
interest that characterize the breakup states. Techniques for determining
other breakup parameters have been successfully developed and applied.
Among them, the density is of particular importance 
because an expansion to low density is 
a basic ingredient of the multifragmentation scenario.

\begin{figure}[ttb]
\centering
    \includegraphics[width=20pc]{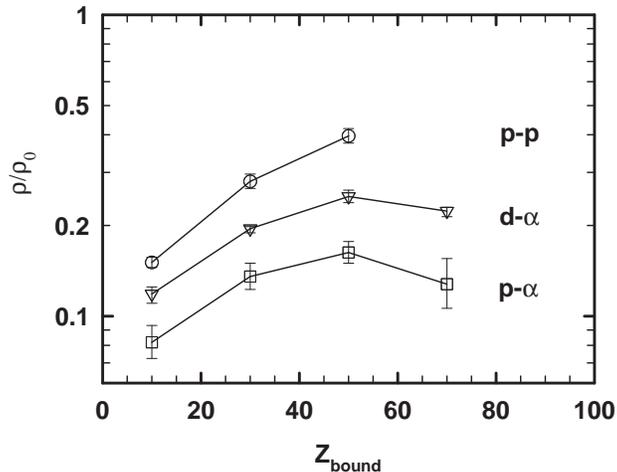} 
\caption{
Breakup density $\rho /\rho_0$ as deduced from 
p-p, p-$^4$He, and d-$^4$He correlation functions measured for
$^{197}$Au on $^{197}$Au reactions at 1000 AMeV. The rapid
decrease of the density with decreasing $Z_{\rm bound}$ reflects the changing
mass of the spectator source. 
Note the logarithmic ordinate scale (from Ref. \protect\cite{fritz99}).
}
\end{figure}

Low densities in agreement with model expectations were recently reported 
for spectator decays following $^{197}$Au on $^{197}$Au reactions at 1000
MeV per nucleon \cite{fritz99}. They were deduced from measured correlation 
functions for proton pairs and for unlike pairs of protons or deuterons 
in coincidence with $\alpha$ particles. 
The correlation functions were found to exhibit the 
surprising property that their variation with $Z_{\rm bound}$ is rather small,
indicating source extensions that do not change dramatically with impact 
parameter in these reactions. 
The observed variation of the density with impact parameter is mainly
caused by the variation of the mass of the intermediate spectator 
system (Fig. 6).

Associated time scales were deduced from the same proton-proton 
data with the technique
of directional analysis \cite{schwarz00}. They are rather short, 
of the order of 20 fm/c, and comparable with the collision time in the 
entrance channel. While this
may indicate that a majority of these protons, selected to have
energies of $E \ge$ 20 MeV in the spectator frame, may originate from early
stages of the collision, it is still obvious that hot fragmentation
initiated by relativistic projectiles is a fast process. The emission
times obtained from fragment-fragment correlations (Fig. 2) are larger 
by only a factor of two.

With experimental values for temperatures and densities at hand, 
the breakup conditions can be displayed 
in the temperature-versus-density plane, commonly
used to characterize the nuclear phase behavior. The three data points
(errors are omitted) shown in Fig. 7 are obtained by correlating the 
measured temperatures $T_{\rm HeLi}$ (Figs. 3 and 5) 
with the p-p densities of Fig. 6. The resulting 
trajectory in the phase diagram tends to higher temperatures and lower 
densities in the more violent collisions. It is important to note that 
this breakup (equilibrium) trajectory is practically
orthogonal to the actual reaction trajectory followed by the system as
it expands from higher to lower temperatures \cite{fried90,papp95}. 

The measured breakup conditions are at 
comparable densities but at temperatures far below the critical point
of nuclear matter as obtained in the calculations of Refs. 
\cite{muell95,sauer76,jaqa83}.
They are also below the critical temperature 
obtained for the finite $^{16}$O
nucleus from the FMD experiment \cite{schnack97}. The conclusion that the 
breakup occurs in the coexistence region thus seems justified also from 
this perspective. Alternatively, it may be derived from the observation
that multifragmentation populates the partition space predicted for the 
coexistence region by the statistical models.

\begin{figure}[ttb]
\centering
    \includegraphics[width=20pc,angle=-90]{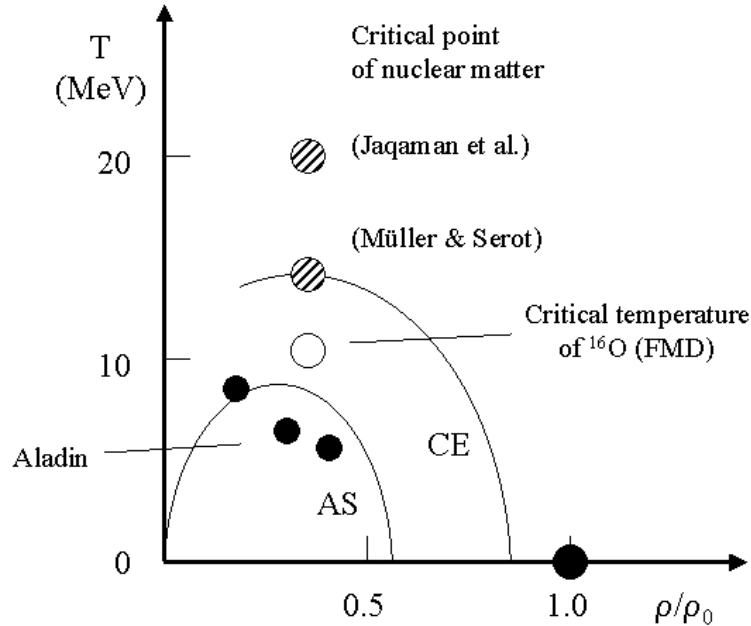}
\caption{Temperature-versus-density diagram with the saturation
point of nuclei (closed circle), critical points of nuclear matter
(hatched, from Refs. \protect\cite{muell95,jaqa83}), the critical
temperature of $^{16}$O according to the FMD \protect\cite{schnack97}, 
and with experimental 
results (dots) obtained by correlating measured temperatures 
and densities. The coexistence
line (CE) and the adiabatic spinodal (AS) for infinite matter,
from Ref. \protect\cite{muell95}, are indicated.
}
\end{figure}

The answer to the question what causes fragmentation is less obvious.
As sketched in the figure, the experimental breakup points are barely 
inside the adiabatic spinodal of the infinite system.
In finite nuclei, this region of volume instability
is probably limited to much lower temperatures. The expanding systems
therefore seem to fragment before coming even close to it. Such a
conclusion may appear speculative at present, but the need
for systematic studies of the instabilities of finite nuclei is rather 
obvious \cite{peth87,colo97a,noe00}.
Surface modes are an alternative to bulk instabilities. While they are 
expected to be slow in homogeneous systems \cite{noe00}, they may be 
rapidly excited during the early stages of the collision. 
Nuclear systems are predicted to be resilient to gentle 
surface excitations but not to major distortions following more violent 
encounters \cite{colo97}.

Additional insight in this direction may come from an improved 
understanding of the kinetic energy spectra of light particles and 
fragments. In cases of bulk fragmentation, the slope temperatures extracted 
from fragment spectra are inconsistent with the chemical temperatures 
obtained with the double-ratio method. However, for spectator decays 
at relativistic bombarding energies, it has been shown that these
temperature values can be 
consistently understood if the slope temperatures are assumed to reflect 
the intrinsic Fermi motion, as assumed in the Goldhaber model \cite{odeh00}. 
Recent calculations with transport models, which incorporate Fermi motion,
support this interpretation \cite{goss97,gait00}.
The energies of spectator fragments are well reproduced, 
and the coexistence of qualitatively
different internal (or local) temperatures and fragment slope
temperatures has been demonstrated (Fig. 8).
The experimental and theoretical findings suggest that fragments are 
preformed at an early stage in these collisions ($\le$ 50 fm/c) before 
they may expand to typical breakup densities. To resolve 
this apparent contradiction 
to the filling of the phase space at breakup, the basis of the statistical 
approaches, is an interesting task for the future.

\begin{figure}[ttb]
\centering
    \includegraphics[width=25pc]{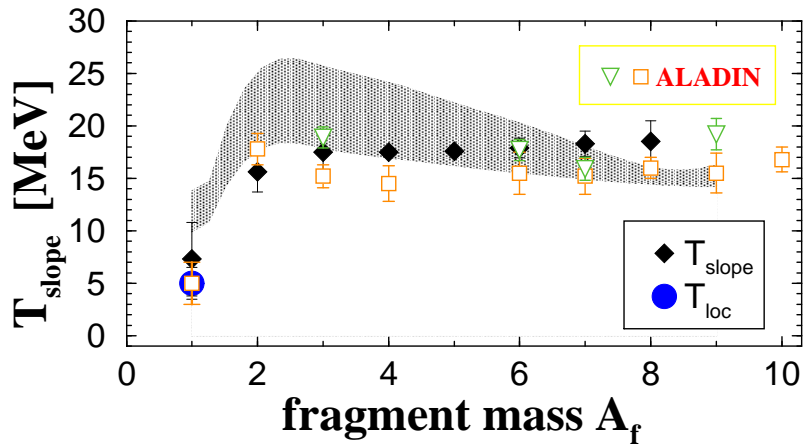} 
\caption{
Comparison of experimental (open symbols, from Ref. \protect\cite{odeh00})
and theoretical (closed diamonds) slope temperatures for spectator decays
as a function of the fragment mass for spectator decays following 
$^{197}$Au on $^{197}$Au collisions. The local nucleon temperature
$T_{\rm loc} \approx$ 5 MeV is indicated at $A_{\rm f}$ = 1, the shaded
area represents the range of slope temperatures obtained by applying 
coalescence to a statistical nucleon distribution
(from Ref. \protect\cite{gait00}).
}
\end{figure}

\section{CONCLUDING REMARKS}

There are many different ways for transforming cold into hot nuclear 
matter, and we have learned to exploit their different features 
in our efforts to understand the fragmentation processes.
The question of how fragments are formed, the problem of identifying the 
dominant mechanisms, has been a recurrent theme during this talk. 
It continues to be a challenge, 
even though remarkable progress has been made within the last few years.

While this requires a realistic modelling of the dynamics, it does not
reduce the statistical approaches in their role and importance. Not only
is it demonstrated that the instabilities governing the fragmentation 
fill the phase space, but 
it is also the statistics and thermodynamics of the process that 
allow us to establish the connection to the nuclear phase transition.

The liquid-gas phase transition continues to act as a major motivation
and has inspired wider investigations of phase transitions in small 
systems, extending beyond the nuclear domain. This more general approach 
seems extremely interesting and rewarding. 
The caloric curve of nuclei has been emphasized in this talk, 
but there are other potential observables, not necessarily less challenging 
on the experimental side, which will be discussed at this conference and
should be further pursued and exploited in the future.  

The uncertainties associated with the measurement of temperatures have been 
extensively investigated. The temperature differences seen
with different methods have led to new insights into the 
fragmentation mechanism.
Experimentally, the bigger challenge probably is the identification of the 
source and the measurement of its excitation energy at 
breakup. There will be limitations in how far we can go with the assumption 
of a well defined breakup configuration.
The continuous evolution of the reaction and emission 
processes may not allow a precise distinction between the equilibrated 
emissions and those preceding it. 

Finally, the majority of the data included here has come from 
elaborate experiments with approximately 4-$\pi$ coverage in the 
respective source 
frames. It is a pleasure to just look at such data, and it is gratifying 
to see the big investments in funds and manpower being justified. Moreover,
it provides encouragement for those who are presently designing or
completing potentially even more powerful detection devices for future
research.

Illuminating and inspiring discussions with my colleagues at the GSI, 
notably A.S.~Botvina, 
H.~Feldmeier, U.~Lynen, W.F.J.~M\"uller, W.~N\"orenberg, W.~Reisdorf 
and C.~Schwarz are gratefully acknowledged. In addition, I 
like to thank L.~Beaulieu, D.~Durand, T.~Gaitanos, J.~{\L}ukasik, 
K.H.~Schmidt and V.E.~Viola for 
providing me with graphics for the talk and for this manuscript.

\end{document}